\documentclass[12pt]{article}

\usepackage{a4wide}
\usepackage{amsfonts}

\newcommand{\nit}{\noindent}

\newcommand{\np}{\newpage}
\newcommand{\dsp}{\displaystyle}
\newcommand{\vs}[1]{\vspace{#1 ex}}
\newcommand{\hs}[1]{\hspace{#1 em}}
\newcommand{\bflr}{\begin{flushright}}
\newcommand{\eflr}{\end{flushright}}
\newcommand{\bc}{\begin{center}}
\newcommand{\ec}{\end{center}}
\newcommand{\ben}{\begin{enumerate}}
\newcommand{\een}{\end{enumerate}}

\newcommand{\be}{\begin{equation}}
\newcommand{\ee}{\end{equation}}
\newcommand{\ba}{\begin{array}}
\newcommand{\ea}{\end{array}}
\newcommand{\ct}{\cite}
\newcommand{\bit}{\bibitem}

\newcommand{\ag}{\alpha}

\newcommand{\gam}{\gamma}
\newcommand{\del}{\delta}
\newcommand{\eps}{\epsilon}

\newcommand{\kg}{\kappa}
\newcommand{\lb}{\lambda}
\newcommand{\sg}{\sigma}

\newcommand{\og}{\omega}
\newcommand{\Gam}{\Gamma}

\newcommand{\Og}{\Omega}
\newcommand{\Lb}{\Lambda}

\newcommand{\lh}{\left(}
\newcommand{\rh}{\right)}

\newcommand{\nb}{\nabla}

\newcommand{\cD}{{\cal D}}

\newcommand{\cH}{{\cal H}}

\newcommand{\der}{\partial}

\begin{document}

\pagestyle{empty} 

\begin{flushright} 
{Nikhef/2017-036}
\end{flushright}
\vs{7}

\bc

{\large \bf D = 1 supergravity and quantum cosmology}
\vs{7}

{\large J.W.\ van Holten}$^a$ 
\vs{4}

\begin{tabular}{l} 
Nikhef, Amsterdam and \\
Lorentz Institute, Leiden University 
\end{tabular} 
\vs{5}
 
\today
\ec
\vs{5} 

\nit
{\small {\bf Abstract} \\
Using a $D = 1$ supergravity framework I construct a super-Friedmann equation for an isotropic 
and homogenous universe including dynamical scalar fields. In the context of quantum theory this 
becomes an equation for a wave-function of the universe of spinorial type, the Wheeler-DeWitt-Dirac 
equation. It is argued that a cosmological constant breaks a certain chiral symmetry of this equation, 
a symmetry in the Hilbert space of universe states, which could protect a small cosmological constant 
from being affected by large quantum corrections.
}
\vfill
 
\footnoterule
\nit 
$^a$ {\em e-mail:} v.holten@nikhef.nl  \\
Contribution to Symphys XVII, Yerevan (Armenia), July 2017

\np
\pagestyle{plain}
\pagenumbering{arabic}

\nit
{\bf 1.\ Introduction} 
\vs{1}

\nit
According to the best available data \ct{planck:2013} the background geometry of our universe 
is spatially flat, the energy density being equal to the critical energy density of which 69.2 \% 
is some kind of dark energy. Although its character and origin are unknown, the simplest account 
of the dark enery is that it represents a cosmological constant, which is however tiny compared to 
any estimate based on fundamental particle physics models; indeed on the Planck scale it is of 
the order of $10^{-123}$ \ct{weinberg:1989}. 

Much effort has gone into trying to establish that such a tiny value follows naturally from 
fundamental physical principles incorporated in quantum field theory and general relativity. 
Weinberg convincingly argued that none of these principles explains the small value of the 
cosmological constant \ct{weinberg:1989}. Therefore a natural explanation requires physical 
principles beyond the standard framework. Supergravity in four or more space-time dimensions 
and string theory are considered candidates for a new starting point. 

In parallel, gradually a new consensus has arisen that in theories allowing a large ensemble 
of possible universes an explanation of a very different kind exists; indeed in a wide landscape 
of universe states there is always a possibility for the creation of a universe that can sustain 
the evolution of complex life, requiring a small cosmological constant to guarantee a long life 
time and the possibility to create a rich set of structures \ct{schellekens:1998, schellekens:2013}. 
Obviously our universe is necessarily of this type. This is the well-known anthropic argument.

Still, anthropic arguments do not preclude the possibility that there exist theories with a 
preference for universes with a small cosmological constant by symmetry arguments, without 
the necessity of fine tuning the parameters. Some promising results in string-inspired 
supergravity have been obtained \ct{kklt:2003, linde:2016}. A survey of different scenarios 
with and without supersymmetry can be found in \ct{nobbenhuis:2004}. 

Assuming cosmology is subject to the laws of quantum theory, if such a natural scenario for a
small cosmological constant exists it should find its expression also at the level of the Hilbert 
space of universe states. In fact it might be easier to detect there than in the field theories 
describing the specific content of our universe. In this paper I present a set of simple models 
of quantum cosmology which have this property. These models extend Friedmann-Lemaitre 
cosmology with a non-standard supersymmetric structure. Whether this structure can be 
incorporated in more conventional supergravity or string scenarios remains to be seen. But 
the mere fact that consistent models exist may encourage looking further in this direction.  
\vs{2}

\nit
{\bf 2.\ Supersymmetric Friedmann-Lemaitre cosmology} 
\vs{1}

\nit
The framework for the models discussed here was presented in \ct{bogers:2015}, based on 
the $D = 1$ supergravity formalism of \ct{jwvh:1995}. This section is devoted to a brief 
summary. The background space-time geometry of an isotropic and homogeneous universe 
is described by the line element 
\be
ds^2 = - N^2(t) dt^2 + a^2(t) g^*_{ij} dx^i dx^j.
\label{2.1}
\ee
Here $a(t)$ is the scale factor, $g^*_{ij}$ is the metric of a constant curvature 3-space, and 
$N(t)$ the lapse function allowing for time reparametrizations. For the metric described by
(\ref{2.1}) the Einstein action including cosmological constant $\Lb$ reduces to 
\be
A_E = \int dt \lh - \frac{3}{N}\, a \dot{a}^2 + N (3k a - \Lb a^3) \rh,
\label{2.2}
\ee
where $k = 0, \pm 1$ is the spatial curvature constant. As a first step we replace the 
scale factor by a variable $\xi^0$ taking values on the full real line, and rescale the lapse
function:
\be
a = e^{\xi^0/\sqrt{6}}, \hs{2} N = n e^{\sqrt{3/2}\, \xi^0},
\label{2.3}
\ee
which allows to rewrite the Einstein action as 
\be
A_E = \int dt\, n \lh - \frac{1}{2n^2}\, \lh \frac{d\xi^0}{dt} \rh^2 + 3k e^{\sqrt{8/3}\, \xi^0}  
 - \Lb e^{\sqrt{6}\, \xi^0} \rh.
\label{2.4}
\ee
In the next step we introduce a set of homogeneous and isotropic scalar fields $\xi^a(t)$,  
$a = 1, ..., r$, on a manifold with metric $G_{ab}[\xi]$ and a scalar potential $V[\xi]$. The full 
action then becomes 
\be
A_G = \int dt\, n \left[ - \frac{1}{2} \lh \cD \xi^0 \rh^2 + \frac{1}{2}\, G_{ab} \cD \xi^a \cD \xi^b 
 - U[\xi^0,\xi^a] \right],
\label{2.5}
\ee
where the full scale-dependent potential is 
\be
U = e^{\sqrt{6}\, \xi^0} V[\xi^a] - 3 k e^{\sqrt{8/3}\, \xi^0},
\label{2.6}
\ee
and we have introduce the reparametrization-invariant derivative
\be
\cD = \frac{1}{n}\, \frac{d}{dt}.
\label{2.7}
\ee
In an obvious short-hand notation this is condensed to 
\be
A_G = \int dt\, n \left[ \frac{1}{2}\, g_{\mu\nu} \cD \xi^{\mu} \cD \xi^{\nu} - U[\xi] \right], \hs{2} 
g_{\mu\nu} = \lh \ba{cc} -1 & 0 \\ 0 & G_{ab} \ea \rh.
\label{2.8}
\ee
So far we have reproduced standard Friedmann-Lemaitre cosmology in a non-standard 
formulation, showing almost full similarity with relativistic particle mechanics. The new step 
taken in \ct{bogers:2015} was to couple this theory to $D = 1$ supergravity. 
The present motivation for doing this is that it gives interesting and apparently consistent 
results. A few comments on the interpretation of the new anti-commuting degrees of freedom 
are offered in the discussion section. 

A multiplet calculus for $D = 1$ supergravity was presented in \ct{jwvh:1995}. Here we use
the gauge multiplet $(n, \chi)$ of lapse function and gravitino, scalar multiplets $(\xi, \psi)$ 
possessing both a commuting and an anticommuting dynamical degree of freedom, and 
fermion multiplets $(\eta, f)$ consisting of a dynamical anticommuting degree of freedom 
and a non-dynamical auxiliary commuting variable. Their transformation rules under 
infinitesimal local supersymmetry transformations with parameter $\eps(t)$ and 
time-reparametrizations with parameter $\ag(t)$ can be cast in the form 
\be 
\ba{ll}
\del n = n \cD \ag - 2i n \eps \chi, & \del \chi = \ag \cD \chi + \cD \eps, \\
 & \\
\del \xi = \ag \cD x - i \eps \psi, & \del \psi = \ag \cD \psi + \lh \cD \xi + i \chi \psi \rh \eps, \\
 & \\
\del \eta = \ag \cD \eta + f \eps, & \del f = \ag \cD f + i \lh \cD \eta - f \chi \rh \eps. 
\ea
\label{2.9}
\ee
These transformations satisfy the closed off-shell commutation algebra 
$\left[ \del_2, \del_1 \right] = \del_3$ with
\be
\eps_3 = \ag_1 \cD \eps_2 - \ag_2 \eps_1, \hs{2}
\ag_3 = 2i \lh \ag_1 \eps_2 - \ag_2 \eps_1 \rh \chi - 2i \eps _1 \eps_2.
\label{2.10}
\ee
To construct a supersymmetric extension of the action $A_G$ we extend all scalar 
degrees of freedom $\xi^{\mu}$ to scalar multiplets, the lapse function is completed to 
a gauge multiplet by a gravitino variable and we introduce two fermion multiplets 
to reconstruct the potential $U$. The result is 
\be
A_{s} = A_{kin} + A_{pot} + A_k, 
\label{2.11}
\ee
where 
\be
\ba{l}
\dsp{ A_{kin} = \int dt\, n \left[ \frac{1}{2}\, g_{\mu\nu} \cD \xi^{\mu} \cD \xi^{\nu} + 
 \frac{i}{2}\, g_{\mu\nu} \psi^{\mu} \lh \cD \psi^{\nu} + \cD \xi^{\lb} \Gam_{\lb\kg}^{\;\;\;\nu} \psi^{\kg} \rh 
 - i g_{\mu\nu} \cD \xi^{\nu} \psi^{\mu} \chi \right],
 }\\
 \\
\dsp{ A_{pot} = \int dt\, n \left[ \frac{i}{2}\, \eta \cD \eta + \frac{1}{2}\, f^2 - W \lh f + i \eta \chi \rh 
 + i \psi^{\mu}\, \der_{\mu} W \eta \right], }\\
 \\
\dsp{ A_k = k \int dt\, n \left[ \frac{i}{2}\, \eta_k \cD \eta_k + \frac{1}{2}\, f_k^2 - \sqrt{6}\, e^{\sqrt{2/3}\, \xi^0} 
 \lh f_k + i \eta_k \chi \rh + 2i e^{\sqrt{2/3}\, \xi^0} \eta_k\, \psi^0 \right]. }
\ea
\label{2.12}
\ee
The function $W[\xi]$ acts as a superpotential; upon eliminating the auxialiary variable $f$ 
we arrive at a potential suitable to describe Friedmann-Lemaitre cosmology provided 
\be
W = e^{\sqrt{3/2}\, \xi^0} \overline{W}[\xi^a] \hs{1} \Rightarrow \hs{1}
\frac{1}{2}\, W^2 = \frac{1}{2}\, e^{\sqrt{6}\, \xi^0}\, \overline{W}^2[\xi^a],
\label{2.13}
\ee
which is clearly non-negative and allows only to consider Minkowski or de Sitter type universes.
Similarly eliminating the auxiliary variable $f_k$ from the last line of (\ref{2.12}) returns the curvature 
term in $U$; the whole expression vanishes for $k = 0$, which is the case mostly discussed in the 
following. 

Note that variation of the action $A_{s}$ with respect to the dynamical degrees of freedom gives
supersymmetric versions of the $D =1$ Klein-Gordon and pseudo-classical Dirac equations, as to 
be expected. Variation with respect to the gauge degrees of freedom produces first-class constraints 
\be
\ba{l}
\dsp{ \frac{1}{2}\, g_{\mu\nu} \cD \xi^{\mu} \cD \xi^{\nu} + \frac{1}{2}\, W^2 + 
  i \eta \lh W \chi + W_{,\mu} \psi^{\mu} \rh }\\
 \\
 \hs{5} \dsp{ +\, k \lh 3 e^{\sqrt{8/3}\, \xi^0} + i e^{\sqrt{2/3}\, \xi^0}
 \eta_k (\sqrt{6}\, \chi - 2 \psi^0) \rh = 0, }\\
  \\
\dsp{ g_{\mu\nu} \cD \xi^{\nu} \psi^{\mu}  + W \eta + k \sqrt{6}\, e^{\sqrt{2/3}\, \xi^0}\, \eta_k = 0. }
\ea
\label{2.14}
\ee
It is to be stressed that these equations are related by supersymmetry and therefore not independent. 
In particular the first Grassmann-even equation follows by a supersymmetry transformation from the
second Grassmann-odd one; thus the second equation plus supersymmetry suffices to reproduce the 
theory. The equations take a more familiar form by introducing a time-reparametrization 
invariant Hubble parameter 
\be
\cH = \frac{1}{\sqrt{6}}\, \cD \xi^0 = \frac{\cD a}{a}.
\label{2.15}
\ee
This brings the constraints to the form 
\be
\ba{l}
\dsp{ 3 \cH^2 = \frac{1}{2}\, G_{ab} \cD \xi^a \cD \xi^b + \frac{1}{2}\, W^2 
 + i \eta \lh W \chi + W_{,\mu} \psi^{\mu} \rh }\\ 
 \\
\dsp{ \hs{5} +\, k \lh 3 e^{\sqrt{8/3}\, \xi^0} + i e^{\sqrt{2/3}\, \xi^0} \eta_k (\sqrt{6}\, \chi - 2 \psi^0) \rh, }\\
 \\
\dsp{ \sqrt{6}\, \cH \psi^0 = G_{ab} \cD \xi^a \psi^b + W \eta + k \sqrt{6}\, e^{\sqrt{2/3}\, \xi^0}\, \eta_k. }
\ea
\label{2.16}
\ee
These are the super-Friedmann equations. Their derivation was the main reason for including 
the non-dynamical gauge degrees of freedom $(n, \chi)$ in the first place. 

Having the full set of equations of motion in hand, it remains to eliminate the non-dynamical 
gauge degrees of freedom; to this end we take the unitary gauge 
\be
N = n e^{\sqrt{3/2}\, \xi^0} = 1, \hs{2} \chi = 0,
\label{2.17}
\ee
in which the line element (\ref{2.1}) takes the standard form in terms of cosmological time $t$.
In this gauge $\cH = a^2 \dot{a} = a^3 H$, with $H$ the usual Hubble parameter in cosmological 
time. In particular the unitary form of the Grassmann-odd Friedmann equation then becomes
\be
\sqrt{6} H \psi^0 = G_{ab} \dot{\xi}^a \psi^b + \overline{W} \eta + k \sqrt{6}\, e^{-\xi^0/\sqrt{6}}\, \eta_k.
\label{2.18}
\ee
As $\overline{W}$ depends only on the scalar fields $\xi^a$, for $k = 0$ the explicit scale-factor 
dependence on the right-hand side through $\xi^0$ disappears completely. The case of flat 
3-space is considered in the following.

\np
\nit
{\bf 3.\ Hamilton formalism}
\vs{1}

\nit
The aim of the hamiltonian formalism is to replace all second-order differential equations of motion 
by pairs of first-order equations of motion. To this end we define the canonical momenta 
\be
p_{\mu} = \frac{\del A_{s}}{\del \dot{\xi}^{\mu}} = g_{\mu\nu} \cD \xi^{\nu} + 
 \frac{i}{2}\, g_{\mu\nu,\lb} \psi^{\nu} \psi^{\lb} - i g_{\mu\nu} \psi^{\nu} \chi,
\label{3.1}
\ee
and rewrite the action in terms of this variable as 
\be
A_{s} = \int dt\, n \left[ p_{\mu} \cD \xi^{\mu} + \frac{i}{2}\, g_{\mu\nu} \psi^{\mu} \cD \psi^{\nu} 
 + \frac{i}{2}\, \eta \cD \eta - H_{s} \right].
\label{3.2}
\ee
The subscript on $H_s$ is to remind the reader this is the hamiltonian for the action $A_s$, 
and not the Hubble parameter. This action agrees with the previous expression (\ref{2.12})
for $k = 0$ if the hamiltonian is
is 
\be
H_{s} = \frac{1}{2}\, g^{\mu\nu} \pi_{\mu} \pi_{\nu} + \frac{1}{2}\, W^2 + i \eta W_{,\mu} \psi^{\mu}
 - i \chi \lh \pi_{\mu} \psi^{\mu} + W \eta \rh,
\label{3.3}
\ee
where we introduced the notation
\be
\pi_{\mu} \equiv g_{\mu\nu} \lh \cD \xi^{\nu} - i \psi^{\nu} \chi \rh = 
 p_{\mu} - \frac{i}{2}\, g_{\mu\nu,\lb} \psi^{\nu} \psi^{\lb}. 
\label{3.4}
\ee
Switching to the non-canonical supercovariant momentum $\pi_{\mu}$ as independent variable the 
equations of motion are reproduced by the hamiltonian upon using the Poisson-Dirac brackets
\be
\ba{ll}
\left\{ \xi^{\mu}, \pi_{\nu} \right\} = \del_{\nu}^{\mu}, & \left\{ \psi^{\mu}, \psi^{\nu} \right\} = - i g^{\mu\nu}, \hs{2}
 \left\{ \eta, \eta \right\} = - i, \\
  & \\
\dsp{ \left\{ \pi_{\mu}, \pi_{\nu} \right\} = - \frac{i}{2}\, \psi^{\kg} \psi^{\lb} R_{\kg\lb\mu\nu}, }& 
\dsp{ \left\{ \psi^{\mu}, \pi_{\nu} \right\} = - \Gam_{\nu\lb}^{\;\;\;\mu} \psi^{\lb}. }
\ea
\label{3.5}
\ee
As the action $A_s$ still contains the gauge variables $(n, \chi)$ the first-class constraints can 
now also be derived in hamiltonian form; they read
\be
H_{s} = 0, \hs{2} Q \equiv \pi_{\mu} \psi^{\mu} + W \eta = 0.
\label{3.6}
\ee
The second constraint reduces the first one to 
\be
H_0 = \frac{1}{2}\, g^{\mu\nu} \pi_{\mu} \pi_{\nu} + \frac{1}{2}\, W^2 + i \eta W_{,\mu} \psi^{\mu} = 0.
\label{3.7}
\ee
It is straightforward to check that $Q$ is  the generator of supersymmetry transformations, and that 
it satisfies the bracket algebra 
\be
\left\{ Q, Q \right\} = - 2 i H_0.
\label{3.8}
\ee
This shows that the fundamental constraint is $Q = 0$, the hamiltonian constraint following 
from this one directly. 

\np
\nit
{\bf 4.\ Quantum cosmology} 
\vs{1} 

\nit
Having constructed a locally supersymmertric pseudo-classical version of the Friedmann-Lemaitre 
equations in hamiltonian form it is straightforward in principle to formulate the corresponding 
quantum theory. However to facilitate the transition from the pseudo-classical Grassman algebra 
generated by $(\psi^{\mu}, \eta)$ to the Clifford algebra for the corresponding quantum operators 
it is convenient to inrtoduce a local Lorentz frame using local frame vectors $e_{\mu}^{\;\,m}$ 
and a spin connection $\og_{\mu\;\,n}^{\;\,m}$ defined by 
\be
g_{\mu\nu} = \eta_{mn} e_{\mu}^{\;\,m} e_{\nu}^{\;\,n}, \hs{2} 
\og_{\mu\;\,n}^{\;\,m} = e^{\nu}_{\;\,n} \nb_{\mu} e_{\nu}^{\;\,m}.
\label{4.1}
\ee
In terms of these one introduces operators 
\be
\sqrt{2}\, \psi^{\mu} e_{\mu}^{\;\,m} \rightarrow \gam^m, \hs{2}
\sqrt{2}\, \eta \rightarrow \ag,
\label{4.2}
\ee
as well as operators for $(\xi^{\mu}, \pi_{\mu})$, whilst the Poisson-Dirac brackets are replaced 
by the (anti-)commutators 
\be
\ba{ll}
\left[ \xi^{\mu}, \pi_{\nu} \right]_- = i \del_{\nu}^{\mu} & \left[ \gam^m, \gam^n \right]_+ = 2 \eta^{mn}
 \hs{2} \left[ \ag, \ag \right]_+ = 2  \\
 & \\
\left[ \gam^m, \pi_{\mu} \right]_- = i \og_{\mu\;\,n}^{\;\,m}\, \gam^n & 
\dsp{ \left[ \pi_{\mu}, \pi_{\nu} \right]_- =  \frac{1}{2}\, \sg^{mn} R_{mn\mu\nu}(\og). }
\ea
\label{4.3}
\ee
Here $R_{mn\mu\nu}(\og)$ is the Riemann tensor re-expressed in terms of the spin connection,
and the $SO(r,1)$ generator $\sg^{mn}$ is constructed as usual in terms of anti-symmetrized 
product of Dirac matrices. The commutation relations for the momentum operator are realized 
in the configuration-space representation in the unitary gauge by 
\be
\pi_{\mu} = - i D_{\mu} = - i \lh \der_{\mu} - \frac{1}{2}\, \og_{\mu mn} \sg^{mn} \rh.
\label{4.4}
\ee
The first-class constraint for the supercharge now is replaced by the 
Wheeler-DeWitt-Dirac equation for a Friedmann-Lemaitre type cosmology
\be
\lh - i \gam^m e^{\mu}_{\;\,m} D_{\mu} + \ag \overline{W} \rh \Psi = 0,
\label{4.5}
\ee
with $\Psi$ a spinorial wave function of the universe. This implies that the quantum-universe 
can possess a polarization in the space of scalar fields. In \ct{bogers:2015} this equation was 
applied to the specific case of the $O(3)$ non-linear $\sg$-model with vanishing potential:
$\overline{W} = 0$. It was shown that in that case the equation has normalizable solutions
and the rate of expansion of the universe represented by the eigenvalues of the operator 
$\pi_0$ becomes quantized.

Here we wish to draw attention to another feature of the Wheeler-DeWitt-Dirac equation, 
which results if we introduce a cosmological constant by the constant superpotential 
\be
\overline{W} = \sqrt{2 \Lb}.
\label{4.6}
\ee
Now observing that 
\be
\ag^2 = 1, \hs{2} \ag \gam^m + \gam^m \ag = 0, 
\label{4.7}
\ee
it follows that $\ag$ has eigenvalues $\pm 1$, and that $\Psi$ can be a pure eigenspinor 
only if the cosmological constant vanishes: $\Lb = 0$; this is completely analogous to chiral 
symmetry in the 4-dimension Dirac equation for quarks and leptons. Therefore $\Lb \neq 0$ 
corresponds to a situation of broken chiral symmetry. In a full quantum-theoretical context 
this chiral symmetry can protect the cosmological constant from becoming large. Of course 
this makes sense only in a multiverse interpretation of quantum cosmology. But from an 
optimistic point of view one could argue that the small observed value of the cosmological 
constant shows that such a mechanism could indeed be at work and that the multiverse 
picture of quantum cosmology ought to be considered seriously.
\vs{2}

\nit
{\bf 5.\ Discussion} 
\vs{1}

\nit
In conclusion the results derived in this paper show that incorporating $D = 1$ supergravity 
in Friedmann-Lemaitre cosmology leads after quantization to a square root of the cosmological 
reduction of the standard Wheeler-DeWitt equation also known as mini-superspace 
\ct{hawking:1983}. The new cosmological quantum equation may be called the 
Wheeler-DeWitt-Dirac equation. This equation has a number of desirable or interesting features: 
it only allows a non-negative cosmological constant; a small value of the cosmological constant
is protected by a type of chiral symmetry; and in certain scenarios the expansion rate of the 
quantum universe is quantized in discrete units. 

Supersymmetric mini-superspace models have been considered before, see e.g.\ 
\ct{death:1988}-\ct{moniz:2010}. These authors mostly consider reductions of $N = 1$, 
$D = 4$ supergravity to Friedmann-Lemaitre or more general Bianchi type universes. The 
construction presented here makes direct use of $D = 1$ supergravity, and to what extent it 
can be derived from or overlaps with the previous approaches is still to be determined. Certainly 
the fermionic components of the $D = 1$ supermultiplets (\ref{2.9}) can not be components of 
$D = 4$ spinors, as this would break local Lorentz invariance. Possibly the anti-commuting 
variables could be some kind of fermionic moduli; or they could be Faddeev-Popov type of ghost 
variables, left over from fixing gauge degrees of freedom in a larger, more complete theory. 
Indeed, although in general supersymmetry is fundamentally different from BRST symmetry, in 
the cosmological context this is no longer obvious as the constraint $Q = 0$ for the supercharge 
and the constraint $\Og = 0$ for the BRST charge have the same form and both operators are 
nilpotent \ct{jwvh:1990}.

While thus presently the origin of the local $D = 1$ supersymmetry remains unclear, the resulting 
quantum cosmology model represented by the Wheeler-DeWitt-Dirac equation is mathematically 
consistent. In fact the construction in terms of $D = 1$ supergravity can be considered conversely 
as a pseudo-classical counterpart of a fully quantum-mechanical square root of the cosmological 
Wheeler-DeWitt equation.  This Wheeler-DeWitt-Dirac equation exhibits the interesting properties 
of quantum cosmology summarized above.
\vs{3}

\nit
{\bf Acknowledgement} \\
The work described here is part of the research programme of Netherlands Foundation of 
Scientific Research Institutes (NWO-I).

\np

\end{document}